\documentclass[pre,superscriptaddress,showpacs,amsmath,amssymb]{revtex4}
\usepackage{graphicx}
\usepackage{graphicx,amsfonts}
\usepackage{epsfig,amsmath}
\usepackage{verbatim}

\begin{document}

\newcommand{\atanh}
{\operatorname{atanh}}
\newcommand{\ArcTan}
{\operatorname{ArcTan}}
\newcommand{\ArcCoth}
{\operatorname{ArcCoth}}
\newcommand{\Erf}
{\operatorname{Erf}}
\newcommand{\Erfi}
{\operatorname{Erfi}}
\newcommand{\Ei}
{\operatorname{Ei}}

\title{Universal Asymptotic Statistics of Maximal Relative Height in
  One-dimensional Solid-on-solid Models}

\author{Gr{\'e}gory Schehr}
\affiliation{Theoretische Physik Universit\"at des Saarlandes
66041 Saarbr\"ucken Germany}
\author{Satya N. Majumdar}
\affiliation{Laboratoire de Physique Th\'eorique et et Mod\`eles
  Statistiques, Universit\'e Paris-Sud, B\^at. 100, 91405 Orsay Cedex,
France}

\date{\today}

\begin{abstract}
We study the probability density function $P(h_m,L)$
of the maximum relative height $h_m$ in a
wide class of one-dimensional solid-on-solid models of
finite size $L$. For all these lattice models, 
in the large $L$ limit, 
a central limit argument shows that, for periodic boundary conditions,
$P(h_m,L)$ 
takes a universal scaling form $P(h_m,L) \sim
(\sqrt{12}w_L)^{-1}f(h_m/(\sqrt{12} w_L))$, with $w_L$ the width of
the fluctuating interface and $f(x)$ the Airy distribution function. 
For one instance of these models, corresponding to the extremely
anisotropic Ising model in two dimensions, this result is obtained by
an exact computation using transfer matrix technique, valid for any
$L>0$. These arguments and exact analytical calculations are
supported by numerical simulations, which show in addition that the subleading 
scaling function is also universal, up to a non universal amplitude,
and simply given by the derivative of the Airy distribution function
$f'(x)$.   
\end{abstract}
\maketitle

\section{Introduction}

The statistics of rare or extreme events in a spatially extended system
such as a fluctuating interface has attracted considerable recent interest.
For example, much theoretical efforts have been devoted to understand
the statistics of first-passage events, both in time~\cite{TempPers}
as well as in  
space~\cite{SpatPers,MDG}
in fluctuating $(1+1)$-dimensional interfaces of the
Edwards-Wilkinson~\cite{EW} 
or the Kardar-Parisi-Zhang~\cite{KPZ} varieties. Some of these
theoretical predictions 
have been recently verified in experimental systems of monatomic 
steps on vicinal surfaces~\cite{Exp}. Another example involves the study of 
extremal Fourier intensities for Gaussian interfaces~\cite{GHPZ}.
In this paper, we focus on the statistics of yet another extreme
observable that has  
generated considerable recent
interest~\cite{RCPS,satya_mrh,satya_mrh2,GK,Lee}, namely the  
statistics of the globally 
maximal relative height 
(MRH) (relative height refers to the height of the interface measured
relative to the 
spatially averaged height in a finite system) in fluctuating interfaces
in its steady state in a finite size system.

The statistics of MRH in fluctuating interfaces has important
theoretical consequences. 
It turns out that
even in cases where the steady state of
the interface may be simple, such as in a Gaussian interface, 
analytical calculation of the probability distribution of the MRH
is nontrivial due to the presence of strong correlations between the heights
at different points of the
interface~\cite{satya_mrh,satya_mrh2}. While the extreme value  
statistics  
of
a set of independent or uncorrelated random variables is relatively easy
to study and well understood along with a complete classification of different
universality classes~\cite{EVS}, much less 
is known when the random variables 
are strongly correlated~\cite{corrEVS}, as in the case of
interfaces. In particular, understanding
the issue of universality of the extreme statistics in such a correlated 
system is a challenging problem to which we focus in this paper. 

One of the simplest examples of a fluctuating interface is provided by the
$(1+1)$-dimensional Edwards-Wilkinson equation~\cite{EW} where the
height $H(x,t)$ evolves via a diffusion equation in presence of an additive thermal noise.
In the long time limit, the system reaches a steady 
state which is
well known~\cite{HZ} to have the Gibbs-Boltzmann form, $P_{st}\propto \exp\left[-\cal 
H\right]$, 
with a simple Gaussian Hamiltonian 
\begin{equation}
{\cal H} = \frac{1}{2}\,\int_0^{L} \left(\frac{\partial H}{\partial x}\right)^2 \, dx 
\label{st1}
\end{equation}
where $H(x)$ is the height of the interface and $L$ is the linear size of the substrate.
Experimental realizations of such a simple Gaussian $(1+1)$-dimensional interface in 
thermal equilibrium 
are provided by 
thermally fluctuating step edges on crystals with attachment/detachment dynamics of 
adatoms~\cite{bart}. For periodic boundary condition (pbc) $H(0)=H(L)$, the process
$H(x)$ defined by the Hamiltonian in Eq. (\ref{st1}) is just a Brownian bridge.
This Hamiltonian has a zero mode since it is invariant under $H(x)\to H(x) +c$ where $c$
is a constant. This is expected since this Hamiltonian does not fix the absolute
height $H$ of the system. Therefore, a more relevant and physically meaningful 
observable is the relative height, $h(x)= H(x)-L^{-1}\int_0^L H(x')dx'$, measured with
respect to the global spatial average. Although the Hamiltonian $\cal H$ in
Eq. (\ref{st1}) retains the same structure in terms of the relative heights, ${\cal H} = 
(1/2)\int_0^L \left({\partial h}/{\partial x}\right)^2 
dx$, the relative heights $h(x)$'s, by definition, satisfy a global constraint,
$\int_0^L h(x)dx =0$. This constraint turns out to play a crucial role in determining the
MRH distribution~\cite{satya_mrh,satya_mrh2}.
 
The variable $h(x)$'s at different space points can
be shown to be strongly correlated. For example, for the pbc, 
the two point correlation function is given by~\cite{satya_mrh2}
\begin{equation}
C(x,L)= \langle h(x_0)h(x_0+x)\rangle = 
\frac{L}{12}\left[1-\frac{6x}{L}\left(1-\frac{x}{L}\right)\right]\,. 
\label{corr1}
\end{equation}
The onsite variance $\langle h^2(x)\rangle = C(0,L)=L/12$ and the 
width $w_L =\sqrt{\langle h^2(x)\rangle}=\sqrt{L/12}$. Thus the typical
relative height scales as $h\sim L^{1/2}$.
Let $h_m$ denote the maximum relative height
in a given sample, {\it i.e.} $h_m={\rm max}[\{h(x)\},\, 0\le x\le L]$. Clearly $h_m$
varies from sample to sample. What is its probability density $P(h_m,L)$?
Since the relative heights $h(x)$ in the steady state are strongly correlated random variables,
the calculation of $P(h_m,L)$ is nontrivial.
In Ref.~\cite{RCPS}, it was demonstrated numerically that $P(h_m,L)$ has a scaling form,
$P(h_m,L)\sim L^{-1/2} f(h_m/\sqrt{L})$. Later, it was proved~\cite{satya_mrh} that
for a continuous interface, this scaling holds for all $L$ (and not just for large $L$).
Expressed in terms of the average width $w_L=\sqrt{L/12}$, 
\begin{equation}
P(h_m,L)= \frac{1}{\sqrt{12} w_L}\, f\left(\frac{h_m}{\sqrt{12} w_L}\right)
\label{scaling1}
\end{equation} 
for all $L$. Moreover, the 
scaling function $f(x)$
was computed exactly~\cite{satya_mrh,satya_mrh2} using path integral techniques and it was shown 
to
be sensitive to the boundary conditions. For example, for the pbc,
it was shown that the Laplace transform of $f(x)$ is given by~\cite{satya_mrh}
\begin{equation}
\int_0^{\infty} f(x) e^{-sx} dx = s\sqrt{2\pi} \sum_{k=1}^{\infty} e^{-\alpha_k 
s^{2/3}2^{-1/3}} ,
\label{lt1}
\end{equation} 
where $\alpha_k$'s are the magnitudes of the zeros of the 
standard Airy function ${\rm Ai}(z)$ on the negative real
axis~\cite{abramowitz}. For example, $\alpha_1=2.3381\dots$, $\alpha_2=4.0879\dots$, 
$\alpha_3=5.5205\dots$ etc ~\cite{abramowitz}. 
Note that the function $f(x)$ defined in Eq. (\ref{lt1}) can be interpreted as a
normalized probability density function (pdf) since
$f(x)\ge 0$ for all $x\ge 0$ and it can be shown that $\int_0^{\infty} f(x) dx=1$.
The function $f(x)$ has the following asymptotic tails~\cite{satya_mrh2}
\begin{eqnarray}
f(x) &\sim & x^{-5}\, e^{- 2\alpha_1^3/{27 x^2}} \quad {\rm as}\quad x\to 0 \nonumber \\
f(x) &\sim & e^{-6 x^2} \quad\quad\quad\quad\quad {\rm as}\quad x\to \infty.
\label{asymfx}
\end{eqnarray}

Remarkably, this function $f(x)$ appeared before in a number of seemingly unrelated problems
in computer science and graph theory and is
known~\cite{FPV,satya_mrh2,BF} as the Airy  
distribution function (not to be 
confused with the Airy function ${\rm Ai}(x)$ itself). Essentially the 
function $f(x)$
describes the pdf of the area under a Brownian excursion on an unit 
interval~\cite{excursion,takacs_invert}. Thus it is interesting that the same pdf also 
describes 
the MRH distribution for a Gaussian interface with pbc~\cite{satya_mrh,satya_mrh2}.
The results 
in Eqs. (\ref{scaling1}) and (\ref{lt1})
provide one of the rare 
exactly solvable cases for the distribution of the extremum of a set of strongly correlated 
random variables. 

A natural question is to what extent this MRH distribution $P(h_m,L)$ and the associated 
scaling function $f(x)$ is universal?
We have already mentioned above that it is sensitive to the boundary condition. This is
natural since $h_m$ is the global maximum over the full sample. However, suppose we
fix the boundary condition to be, say periodic, and ask how sensitive $P(h_m,L)$ is
on the details of the short range interaction in the Hamiltonian $\cal H$ in Eq. (\ref{st1})
in the steady state? In this paper we address this issue and study, both analytically 
and numerically, the MRH distribution in a class of one dimensional 
solid-on-solid (SOS) models~\cite{Forgacs}. The SOS model is defined on a lattice with 
$L$ sites labelled $0,1,\ldots L-1$,  
with pbc (so that the site $L$ is identified with site $0$). 
Note that we have set the lattice constant to be unity so that $L$ also
represents the total length of the substrate as in the continuous Hamiltonian
in Eq. (\ref{st1}). 
The model,
at equilibrium, is described by the Hamiltonian 
\begin{eqnarray}
{\cal H}_p = K\sum_{i=0}^{L-1} |H_i - H_{i+1}|^p \label{def_hamiltonian}
\end{eqnarray}
where $H_i \in  ]-\infty, +\infty[$ is a continuous height variable and 
$p>0$ is a positive parameter. This model is a continuous-height version
of the discrete-height SOS models where the height variables $H_i$'s are 
integers~\cite{Forgacs}.
For the discrete-height SOS model, one can generate a family of models by tuning the
parameter $p$. For example, the case $p=1$ corresponds to the extremely anisotropic Ising 
model
in $2$ dimensions, where as the case $p\to \infty$ corresponds to the restricted
SOS model where neighbouring heights can differ at most by one unit~\cite{Forgacs}. 
Several authors have studied the $p=1$ model, both for discrete heights~\cite{discrete}
and for continuous heights~\cite{burkhardt},   
in the context of wetting phenomena in two dimensions.

In this paper we restrict ourselves, for simplicity, to the model in Eq. (\ref{def_hamiltonian}) with
continuous height $H_i$'s, though our main asymptotic results will be valid even for 
the discrete-height models. We define the 
relative heights $h_i= H_i- \sum_i H_i/L$
and study the distribution of the MRH $P(h_m, L)$ in
this model. Note that the case $p=2$ in Eq. (\ref{def_hamiltonian}) corrresponds 
to the spatially discretized version of the continuous-space Gaussian interface model
in Eq. (\ref{st1}). Therefore, in this case, one would expect that as the number of sites
$L$ becomes asymptotically large, one would recover the continuum limit results for 
the MRH
in Eqs. (\ref{scaling1}) and (\ref{lt1}). Less obvious is what happens when $p\ne 0$.
Our main 
object here is to study the
MRH distribution for arbitrary $p>0$, to see how it depends on the parameter $p$
and also to investigate the leading finite size effects.

Our main results are summarized as follows.  
We will demonstrate that asymptotically for large $L$, the MRH distribution
$P(h_m,L)$ is described by the scaling form in Eq. (\ref{scaling1}) where
the width $w_L$ depends on $p$ (weakly) but the scaling function $f(x)$ is 
universal, {\it i.e.} independent of $p$ and is described by the Airy
distribution  
function in Eq. (\ref{lt1}). This result is less obvious apriori for $p\ne 2$.
We will first present a general argument based on the central limit
theorem that 
will suggest this universality with respect to $p$. Next we will
present numerical  
simulations in support of this universal result. In fact, this central limit
argument holds even for a more general class of SOS models defined by the Hamiltonian
\begin{equation}
\exp[-{\cal H}] \propto \prod_i g\left(|H_{i}-H_{i+1}|\right)
\label{Hamilg}
\end{equation}
where $g(x)$ is an arbitrary positive, symmetric and normalized (to
unity) function, but with a  
finite  second 
moment $\sigma^2= \int_{-\infty}^{\infty} g(x) x^2 dx$. 

For the special case $p=1$ of the model in Eq. (\ref{def_hamiltonian}), we will present
an exact calculation of $P(h_m,L)$ for any $L$ that proves explicitly this universality
in the large $L$ limit.
Moreover, for $p=1$, we also calculate exactly the subleading correction 
to the leading scaling form for large $L$ and then demonstrate numerically that even
the subleading scaling function, up to a $p$ dependent amplitude, is also
universal with respect to $p$ and the associated subleading scaling function   
is simply the derivative $f'(x)$ of the Airy distribution function in Eq. (\ref{lt1}).    

The paper is organized as follows. In the next section (II), 
we present an argument for the universality of the asymptotic MRH distribution based on the 
central limit theorem.
In section III,
we present a general
set-up to compute the MRH distribution in the general SOS models defined in Eq. (\ref{Hamilg})
for arbitrary $L$.
Section IV contains an exact result, valid for all $L$, for the $p=1$ case.
In section V, we present the details of the numerical simulations. We conclude
in section VI with a summary and outlook for future studies. Finally, in appendix A, we
present an exact calculation of the width of the interface defined in Eq. (\ref{Hamilg}).

\section{A General Universality Argument based on Central Limit Theorem}

Our starting point is the general Hamiltonian in Eq. (\ref{Hamilg}).
The product form on the right hand side of Eq. (\ref{Hamilg}) indicates
that one can interpret the successive height differences $H_{i+1}-H_i=\xi_i$'s to be
a set of random numbers drawn from the joint distribution
\begin{equation}
P\left[\{\xi_i\}\right]= N_L\, g(\xi_1)\,g(\xi_2)\,\ldots g(\xi_L)\, \delta\left(\sum_{i=1}^L
\xi_i\right)
\label{jointl}
\end{equation}
where $N_L$ is such that the joint distribution is normalized and $g(x)$ is
a symmetric, normalized pdf with a finite second moment
$\sigma^2$, but otherwise arbitrary. The delta
function ensures that the path is periodic, $H_0=H_L$.
Therefore, the height profile $H_i$ can be interpreted as a discrete random `bridge'
process, bridging the two ends $H_0=H_L$ and in between, performing a discrete
random walk by choosing at each step a random jump variable $\xi$ drawn independently
from the normalized pdf ${\rm Prob}(\xi=x)=g(x)$ with a finite variance $\sigma^2$.
For the special case of the Hamiltonian in Eq. (\ref{def_hamiltonian}) the
noise pdf is
\begin{equation}
{\rm Prob}(\xi=x)=g_p(x) = \frac{1}{B_p} e^{-K |x|^p}
\label{noisedist}
\end{equation}
where the normalization constant $B_p = \int_{-\infty}^{\infty} dx\, e^{-K |x|^p}$ 
and the variance $\sigma_p^2=\int_{-\infty}^{\infty}x^2\, g_p(x)\, dx$
are obtained as
\begin{equation}
B_p=\frac{2\Gamma(1/p)}{p}\, K^{-1/p}\quad;\quad \quad 
\sigma_p^2=\frac{\Gamma(3/p)}{\Gamma(1/p)}\,K^{-2/p}\,.
\label{norm0}
\end{equation}
In particular, $B_1=2/K$ and $\sigma_1^2= 2/K^2$ will be used later in section IV.

In the limit of large $L$, a standard central limit argument asserts that the
discrete random bridge process $H_i$ will converge in law, up to a nonuniversal
scale factor that depends on the variance $\sigma^2$ of the noise pdf $g(x)$, to the 
continuous Brownian bridge process $H(x)$ described by the Hamiltonian in Eq. (\ref{st1}).
Of course, this will be true as long as ${\rm Prob}(\xi=x)=g(x)$ decays faster than
$|x|^{-3}$ for large $|x|$, {\it i.e.} the second moment $\sigma^2$ of $g(x)$ is finite.
Naturally the same conclusion will hold for the relative height $h_i=H_i-\sum_i H_i/L$
as well. More precisely, for large $h_i$ and large $L$, the dimensionless
scaled variable $h_i/w_L$ (where $w_L$ is the width defined via the relation
$w_L^2= \sum_i \langle h_i^2\rangle/L$) will have identical asymptotic 
statistics as that of its continuous counterpart $h(x)/\sqrt{L/12}$. 
Naturally the scaled MRH $h_m/w_L$ where $h_m= {\rm max}\left(\{h_i\}, i=0,1,\cdots ,L-1\right)$
is expected to follow the same asymptotic statistics as that of its continuous counterpart,
as long as $\sigma^2$ is finite.
This argument thus predicts that for arbitrary symmetric $g(x)$ with a finite variance
$\sigma^2$, the asymptotic MRH distribution will converge to
\begin{equation}
P(h_m,L) \to \frac{1}{\sqrt{12} w_L}\, f\left(\frac{h_m}{\sqrt{12} w_L}\right)   
\label{univd}
\end{equation}
where $f(x)$ is the Airy distribution function defined in Eq. (\ref{lt1}). 

Indeed, `nonuniversality' of the MRH distribution, {\it i.e.} its dependence on the noise pdf ${\rm 
Prob}(\xi=x)=g(x)$ 
enters only through the width $w_L$ in Eq. (\ref{univd}).
In fact, for 
the discrete process with arbitrary noise pdf $g(x)$, one can
calculate the width $w_L$ exactly (as derived in appendix A) for all L and one gets
\begin{eqnarray}
w^2_L = \frac{1}{L}  \sum_i \langle h_i^2 \rangle =\frac{L^2-1}{12}\Delta_L \quad, \quad 
\Delta_L=\frac{\int_{-\infty}^{\infty} [{\tilde g}(k)]^{L-2}\,[{\tilde g}'(k)]^2\, dk}
{\int_{-\infty}^{\infty} [{\tilde g}(k)]^{L}\, dk}
\label{width}
\end{eqnarray}
where ${\tilde g}(k)=\int_{-\infty}^{\infty} g(x) e^{ikx} dx$ is the Fourier transform 
of $g(x)$ and ${\tilde g}'(k)= d{\tilde g}/dk$. In particular, for large $L$, one can
show (see appendix A) 
that $\Delta_L \approx \sigma^2/L$ where $\sigma^2=-{\tilde g}''(0)$ is the
variance of $g(x)$. Then
$w_L \approx \sigma \sqrt{L/12}$, {\it i.e.} the same result as in the continuous case, apart
from the nonuniversal scale factor $\sigma$.

This completes our argument for the universality of the MRH distribution
in the SOS model defined in Eq. (\ref{Hamilg}) which predicts that,
for large $L$, the MRH distribution for arbitrary symmetric $g(x)\ge 0$ (with a
finite second moment $\sigma^2$)
would have the scaling form in Eq. (\ref{univd}) with $w_L\approx 
\sigma \sqrt{L/12}$ and with a universal scaling function $f(x)$ (Airy distribution 
function) as defined in Eq. (\ref{lt1}). The universal asymptotic scaling function predicted in this 
section are further
supported by the exact result in section IV for the $p=1$ case of the model in Eq. 
(\ref{def_hamiltonian}) and numerical simulations for other values of
$p$ as presented in section V. 

\section{The General Set-up to Compute the MRH Distribution for Arbitrary $L$} 

Let us outline the general set-up to calculate the MRH distribution in the SOS
models defined by the Hamiltonian in Eq. (\ref{Hamilg}). 
This set-up is a discrete analogue of the method used for the continuous Gaussian interface 
model~\cite{satya_mrh,satya_mrh2}.
In terms of the relative heights
$h_i= H_i-\sum_i H_i/L$, the SOS Hamiltonian retains the same form as in 
Eq. (\ref{Hamilg}). 
However, the relative height $h_i$'s now, by definition, satisfy
a global constraint, $\sum_i h_i=0$. Moreover, the pbc requires
$h_0=h_L$. Thus, the joint probability distribution of the relative heights $\{h_i\}$
with $i=0,1,\ldots, L-1$ 
in the steady state can be written as
\begin{equation}
P\left[\{h_i\}\right]= A_L\, \left[\prod_{i=0}^{L-1} g\left(|h_i-h_{i+1}|\right)\right]\,  
\delta\left[\sum_{i=0}^{L-1} h_i\right]
\label{jp1}
\end{equation}
where the delta function on the right hand side ensures the global constraint $\sum_i h_i=0$. The 
normalization
constant $A_L$, to be calculated later, is fixed by demanding that the joint distribution
$P[\{h_i\}]$ is normalized to unity.

We wish to compute the pdf $P(h_m, L)$ of the MRH $h_m= {\rm max}\left[h_0, h_1, \ldots, 
h_{L-1}\right]$. It turns out to be easier first to compute the cumulative distribution
$F(h_m,L) = {\rm Prob}\left[{\rm max}(\{h_i\})< h_m, L\right]$. The pdf of the MRH
is simply the derivative, $P(h_m,L)= \partial F(h_m,L)/{\partial h_m}$. Note that
$F(h_m,L)$ is the probability that the maximum of all heights are less than $h_m$,
which is the same as the probability that all the height variables $h_i\le h_m$. Using
the joint probability distribution of $h_i$'s in Eq. (\ref{jp1}), one can then
express $F(h_m,L)$ as a multiple integral
\begin{eqnarray}
F(h_m, L) &= & \int_{-\infty}^{h_m}dh_0\, \int_{-\infty}^{h_m} dh_1\ldots 
\int_{-\infty}^{h_m} dh_{L-1}\, P\left[\{h_i\}\right] \nonumber \\
&=& A_L \int_{-\infty}^{h_m} {\cal
  D}h \left[\prod_i g\left(|h_i-h_{i+1}|\right)\right] 
\delta\left(\sum_i h_i \right) 
  \label{starting_formula}
\end{eqnarray}
where ${\cal D}h$ is a shorthand notation for ${\cal D} h= dh_0\, dh_{1}\ldots dh_{L-1}$.
Making a change of variable $y_i= h_m-h_i$, Eq. (\ref{starting_formula}) becomes
\begin{equation}
F(h_m,L) = A_L \int_0^{\infty} {\cal D} y 
\left[\prod_i g\left(|y_i-y_{i+1}|\right)\right]
\delta\left(\sum_i y_i -h_m L\right)\,.
\label{stfm2}
\end{equation}
Note that the $h_m$ appears on the right hand side only inside the delta function
as a combination $A=h_m L$. 
Furthermore, $h_m$ can not be negative as this would indicate that all the
relative heights $h_i\le 0$ which can not be true due to the exact constraint
$\sum_i h_i=0$. Thus, the pdf $P(h_m,L)$ has nonzero support only for $0\le h_m \le \infty$.
Writing $F(h_m,L)= {\cal F} (A,L)$ and taking its Laplace transform 
with respect to $A$ we get
\begin{eqnarray}
\int_0^\infty {\cal F}(A,L) e^{-\lambda A}dA = A_L \, Z(\lambda,L) \quad ;
\quad
Z(\lambda, L)= \int_0^{\infty} {\cal D}y\, \prod_i g\left(|y_i-y_{i+1}|\right)\, e^{-\lambda y_i}\,.
 \label{pfg}
\end{eqnarray}
For the special case when $g(x)\propto \exp[-K|x|^p]$ as in Eq. (\ref{def_hamiltonian}),
the partition function is given by
\begin{equation}
Z(\lambda,L) = 
\int_0^\infty {\cal D} y
\exp{\left(-K\sum_i |y_i -
  y_{i+1}|^p - \lambda \sum_i y_i\right)}\,. 
\label{starting_laplace}
\end{equation}

Thus if we can determine the partition function $Z(\lambda, L)$, then by inverting
the Laplace transform we get $F(h_m,L)$. The normalization constant $A_L$ can then be fixed
by using the fact that $F(h_m\to \infty, L)=1$ (which follows from the fact that the MRH
pdf must be normalized to unity). In fact, this procedure can be directly implemented
in Eq. (\ref{pfg}) by taking the $\lambda\to 0$ limit.
As $\lambda\to 0$, $\int_0^\infty {\cal F}(A,L) e^{-\lambda A}\, dA\to 1/\lambda$
using the fact $F(A\to \infty,L)=1$. Then it follows that
\begin{equation}
A_L = \lim_{\lambda\to 0} \frac{1}{\lambda Z(\lambda,L)}\,.
\label{norm1}
\end{equation} 
The partition function $Z(\lambda, L)$ in Eq. (\ref{pfg}) can, in principle,
be computed by a transfer matrix method for a general $g(x)$. However, an explicit exact solution 
for all $L$ can be obtained for the $p=1$ case in Eq. (\ref{starting_laplace}), as demonstrated 
in the next section.

We end this section by pointing out an interesting connection between the MRH distribution
in Eq. (\ref{starting_formula}) and a first-passage or no barrier crossing probability
of a discrete random acceleration process. It is evident from Eq. (\ref{jp1}) that
the relative height variables $h_i$ also form a random bridge (pbc) as the absolute 
height $H_i$'s, except with an additional long range constraint $\sum_i h_i=0$
as manifest by the delta function in the joint distribution in Eq. (\ref{jp1}). 
Thus the process $h_i$, though locally a random walk, it is conditioned
to return to its initial point (pbc) and also has to remember that at the end of $L$ steps the 
total `area' under
the walk must be identically $0$. This long range global constraint thus induces a
`memory' effect and makes the evolution of $h_i$ a non-Markovian process. This fact was
already noted in the context of the continuous Gaussian interface model~\cite{satya_mrh2,MDG}.
This non-Markovian process can be recast as a Markov process via the following trick.
We define an additional `area' variable
$A_i = \sum_{j=0}^{i-1} h_{i-1}$ with $A_0=0$. Then the joint two variable process
$(h_i,A_i)$ actually evolves by a Markov process
\begin{eqnarray}
h_{i+1} = h_i + \xi_i \quad, \quad A_{i+1} = A_i + h_i \label{discr_rnd_acc}
\end{eqnarray}
where $\xi_i$'s are drawn from the joint distribution in Eq. (\ref{jointl}).
 
Thus, Eq. (\ref{discr_rnd_acc}) is just the discrete version of
the random acceleration problem~\cite{sinai,burkhardt_jointprob}
({\it i.e.} the second derivative of $A$ is noise).
This joint process $(h_i,A_i)$ starts with its initial value
$(h_0,A_0=0)$ at step $i=0$ and ends at its final value $(h_L=h_0,A_L=0)$ after
$L$ steps ensuring both the pbc and the
zero area constraint at the end of the process, $A_L=\sum_{i=0}^{L-1} h_i=0$.
This then completely describes the inferface height profile $h_i$ at equilibrium.
The cumulative MRH distribution
$F(h_m,L)$ in Eq. (\ref{starting_formula}) is then
just the probability that the joint process $(h_i,A_i)$ evolving via
Eq. (\ref{discr_rnd_acc}) reaches from its initial value $(h_0,A_0=0)$ to its final value
$(h_L=h_0, A_L=0)$, but with the restriction that the $h_i$'s stay below the level $h_m$
up to $L$ steps. This is then just a restricted propagator of the discrete random
acceleration process.

\section{Exact Solution for arbitrary $L$ using Tansfer matrix for $p=1$.}

The partition function 
$Z(\lambda,L)$ in Eq. (\ref{pfg}) can in principle be computed, for arbitrary $g(x)$, using 
a standard transfer
matrix technique. The transfer matrix $T$ associated to $Z(\lambda,L)$ 
in
Eq. (\ref{pfg}) is defined via 
\begin{eqnarray}
\langle x | T | y \rangle = \exp{(-\lambda x/2)}\,g(|x-y|)\, \exp{(-\lambda y/2)}
\end{eqnarray}
and the partition function is simply the trace, $Z(\lambda,L) ={\rm Trace}[{T}^L]$.
Let us define $\phi(x)$ as its eigenfunctions that satisfy the eigenvalue equation
\begin{equation}
\int_0^{\infty} dy\, \langle x|T|y\rangle\, \phi(y)= E \phi(x)
\label{eigen1}
\end{equation} 
where $E$ label the eigenvalues.
Making a substitution, $\phi(x) =
\psi(x)\, e^{-\lambda x/2}$ one gets an integral equation 
\begin{eqnarray}
\int_0^\infty dy\, g(|x-y|)\,\exp{(-\lambda y)}\, \psi(y) = E
\psi(x) \label{transfer_eq}
\end{eqnarray}

For an arbitrary $g(x)$, the integral equation (\ref{transfer_eq}) is hard to solve.
However, one can make progress for the special case of the Hamiltonian in Eq. 
(\ref{def_hamiltonian}) with $p=1$ that corresponds to $g(x)=\exp[-K|x|]$ 
to which we now focus.
For this case, the integral equation 
(\ref{transfer_eq}) can be transformed into a Schr\"odinger-like (though not quite the 
same) differential equation
by using the identity $(-d^2/dx^2 +K^2) e^{-K|x-y|}= 2K \delta(x-y)$ ~\cite{burkhardt}.
We get
\begin{eqnarray}
\psi''(x) + \left(\frac{2K}{E} e^{-\lambda x} - K^2\right) \psi(x) = 0
\label{transfer_schrodinger} 
\end{eqnarray}
with the boundary conditions
\begin{eqnarray}
(i) \lim_{x \to \infty} \psi(x) = 0 \quad, \quad (ii)
\frac{\psi'(0)}{\psi(0)} = K \label{transfer_pbc}
\end{eqnarray}
which follow from Eq. (\ref{transfer_eq}). 

The differential equation (\ref{transfer_schrodinger}) can be further reduced
to a standard Bessel form by a change of variable.  
The general solution to this equation has two
linearly independent parts, but 
one of them is not allowed due to the condition $(i)$ in
Eq. (\ref{transfer_pbc}).  One finally gets the solution of
Eq. (\ref{transfer_schrodinger}) as
$\psi(x) = A\, J_{{2K}/{\lambda}}[ ({2}/{\lambda})(\sqrt{2K/E})
  e^{-\lambda x/2}]$, 
where $J_\nu(x)$ is a Bessel function and
$A$ is an arbitrary amplitude. The 2nd boundary condition $(ii)$ in
Eq. (\ref{transfer_pbc}) determines the eigenvalue $E$ as a root
of the following equation
\begin{eqnarray}
J'_{2K/\lambda}(u)/J_{2K/\lambda}(u) = - 2K/(\lambda u) \label{pbc_2}
\end{eqnarray}
where $u= 2\sqrt{2K/E}/\lambda$. Using the identity $J'_{\nu}(x) =
J_{\nu-1}(x) - (\nu/x) J_{\nu}(x)$, Eq. (\ref{pbc_2}) becomes:
\begin{eqnarray}
J_{2K/\lambda - 1}(u) = 0 \quad, \quad i.e. \quad E_n = \frac{2}{K}
\left(\frac{\lambda}{2K} j_{2K/\lambda - 1,n} \right)^{-2}
\end{eqnarray} 
where $j_{\nu,n}$ is the $n^{\text{th}}$ root of $J_\nu(x)$ on the real
axis. Thus $Z(\lambda,L)$ is given by the exact formula, valid for all $L$,
\begin{eqnarray}
Z(\lambda,L) = B_1^L \sum_n
\left(\frac{\lambda}{2K} j_{2K/\lambda - 1,n} \right)^{-2L} 
\end{eqnarray}
where $B_1=2/K$. This completes our derivation of the exact partition function
$Z(\lambda, L)$ for all $L$. 
\vspace{0.4cm}

\noindent {\bf Asymptotic results for large $L$:}
We next focus on the large $L$ limit. If one na\"ively takes the large
$L$ limit 
of $Z(\lambda,L)$ keeping $\lambda$ fixed, it is just enough to
retain only the
largest eigenvalue of the transfer matrix in the partition sum.
However to extract the leading scaling behavior of the cumulative distribution $F(h_m,L)$,
we need to take the limits $L\to \infty$, $h_m\to \infty$ simlutaneously while keeping the ratio
$h_m/\sqrt{L}$ fixed. Since $A=h_m\,L \sim L^{3/2}$ is the variable conjugate to the Laplace 
variable
$\lambda$ in $Z(\lambda, L)$, the appropriate scaling limit is
$L\to \infty$, $\lambda\to 0$ but keeping the scaling combination 
$\lambda L^{3/2}$ fixed. 
Hence we define a new variable
\begin{equation}
s= \sigma_1 \lambda L^{3/2}; \quad \quad {\rm where}\quad \sigma_1=\frac{\sqrt{2}}{K}.
\label{defs}
\end{equation}
We next expand
$\left(\frac{\lambda}{2K} j_{2K/\lambda - 1,n} \right)^{-2L}$ at large
$L$, keeping $s$ fixed. To proceed, we use the expansion of
$j_{\nu,n}$ for large $\nu$, $n$ fixed \cite{abramowitz}:
\begin{eqnarray}
j_{\nu,n} = \nu + \frac{\alpha_n}{2^{1/3}} \nu^{1/3} + {\cal
  O}(\nu^{-1/3}) \label{expansion_bessel_root} 
\end{eqnarray}
where $\alpha_n$ is the amplitude of $n^{\text{th}}$ zero of the Airy
function ${\rm Ai}(z)$ on 
the negative real axis. Using this expansion
(\ref{expansion_bessel_root}), we obtain 
\begin{eqnarray}
Z(\lambda,L) = B_1^L\, \sum_n
e^{-\alpha_n s^{2/3} 2^{-1/3}}\left(1 + \frac{s}{\sqrt{2L}} + {\cal
O}(L^{-1})  \right)\,.
\label{Z_L_final}
\end{eqnarray}
Substituting this result in Eq. (\ref{starting_laplace}), one gets for $p=1$
\begin{eqnarray}
\int_0^\infty {\cal F}(A,L)\, e^{-\lambda A}\,dA = A_L\, B_1^L\,   
\sum_n
e^{-\alpha_n s^{2/3} 2^{-1/3}}\left(1 + \frac{s}{\sqrt{2L}} + {\cal
  O}(L^{-1}) \right)
\end{eqnarray}
so that, using the Bromwich formula for Laplace inversion and substituting $A=h_m L$ one obtains
the cumulative distribution as
\begin{equation}
F(h_m,L) = A_L\, B_1^L\, \int_{\lambda_0-i\infty}^{\lambda_0+i\infty}
  \frac{d\lambda}{2\pi i}\, e^{\lambda h_m L}\, \sum_n
e^{-\alpha_n (\sigma_1 \lambda L^{3/2})^{2/3} 2^{-1/3}}\left(1 +
  \frac{\sigma_1 \lambda L^{3/2}}{\sqrt{2L}} + {\cal 
  O}(L^{-1}) \right) \label{last_expr_F}
\end{equation}
where the integration is along any imaginary axis whose real part
$\lambda_0$ must be to the right of all the singularities of the
integrand. Performing the change of variable
$s = \sigma_1 \lambda L^{3/2}$ and taking the derivatives with respect
to $h_m$ one finds that $P(h_m,L)$ has the structure
\begin{eqnarray}
P(h_m,L) = \frac{A_L\, B_1^L}{\sqrt{2\pi}\, \sigma_1 L^{3/2}}\,
\left[\frac{1}{\sigma_1 L^{1/2}} f\left(\frac{h_m}{\sigma_1 L^{1/2}}\right) + 
  \frac{1}{\sigma_1 L}f_1\left(\frac{h_m}{\sigma_1 L^{1/2}}\right) +
  \frac{1}{\sigma_1 L^{3/2}}f_2\left(\frac{h_m}{\sigma_1 L^{1/2}}\right) +
  ...\right] \label{exp_P} 
\end{eqnarray}
where $f(x)$ is the leading scaling function in the large $L$
limit. From Eq. (\ref{last_expr_F}), its Laplace transform $\hat f(s)$
is given by
\begin{eqnarray}
\hat f(s) = \int_0^\infty dx\, f(x)\,e^{-sx} = s\sqrt{2\pi}\sum_n
e^{-\alpha_n s^{2/3} 2^{-1/3}}
\label{f}
\end{eqnarray} 
as in Eq. (\ref{lt1}) and thus $f(x)$ is the Airy-distribution function.
Similarly it is easy to check from Eq. (\ref{last_expr_F}) that 
\begin{equation}
{\hat f_1(s)}=\int_0^{\infty} dx\, f_1(x)\, e^{-sx} = \frac{1}{\sqrt{2}}\,s\, {\hat f(s)}\,.
\label{f1}
\end{equation}

We are now ready to determine the normalization constant $A_L$. Since the pdf
$P(h_m,L)$ is normalized to unity, we have $\int_0^{\infty} P(h_m,L)\, dh_m=1$.
Integrating Eq. (\ref{exp_P}) and using the facts: (i)  $\int_0^{\infty} f(x)\,dx={\hat f(0)}=1$
(which can be shown from Eq. (\ref{f})) 
and (ii) $\int_0^{\infty} f_1(x)\,dx = {\hat f_1(0)}=0$ (as follows from Eq. (\ref{f1})),
we get
\begin{equation}
\frac{A_L \, B_1^L}{\sqrt{2\pi}\, \sigma_1 L^{3/2}}[1+ {\cal O}(L^{-1})]=1
\label{norm2}
\end{equation}
that gives, using $B_1=2/K$,
\begin{equation}
A_L = \sqrt{2\pi}\, \sigma_1\, L^{3/2}\, \left(\frac{K}{2}\right)^L\, \left[1 + {\cal 
O}(L^{-1})\right]\,.
\label{ampl}
\end{equation}  
Substitution in Eq. (\ref{exp_P}) then gives the asymptotic scaling function
and its first corrrection
\begin{equation}
P(h_m,L) = \frac{1}{\sigma_1 L^{1/2}}
f\left(\frac{h_m}{\sigma_1
  L^{1/2}}\right) +
  \frac{1}{\sigma_1 L}f_1\left(\frac{h_m}{\sigma_1 L^{1/2}}\right) + ..
\label{final1}
\end{equation}
where $f(x)$ and $f_1(x)$ are given by their Laplace transforms in Eqs. (\ref{f}) and (\ref{f1})
respectively. This is our main result of this section.
 
It turns out that the Laplace transform in Eq. (\ref{f}) can be formally 
inverted~\cite{takacs_invert,satya_mrh2}
\begin{eqnarray}
f(x) = \frac{2\sqrt{6}}{x^{10/3}}\sum_{n=1}^\infty e^{-b_n/x^2}
b_n^{2/3} U(-5/6,4/3,b_n/x^2) 
\label{Airy_dist}
\end{eqnarray}
where $b_n = 2\alpha_n^3/27$ and $U(a,b,z)$ is the confluent
hypergeometric function \cite{abramowitz}. This inversion is useful as it can then be 
evaluated using Mathematica and subsequently one can compare it 
with the numerical simulations later.
The leading `correction to scaling' function $f_1(x)$ has received some
attentions recently in a class of random walk problems that arise in
the context of rectangle packing problem
\cite{flajolet_correc_scal,CM} as well as in the problem of
computing the total flux of diffusing particles to a spherical trap in $3$ dimensions~\cite{CMZ}. 
It is interesting that in the present problem we can compute this
correction to scaling function $f_1(x)$ exactly. Inverting the Laplace transform
in Eq. (\ref{f1}), we get $f_1(x)=f'(x)/\sqrt{2}$ where $f'(x)$ can be evaluated by
taking the derivative of Eq. (\ref{Airy_dist}) and using properties of $U(a,b,z)$.
This gives
\begin{eqnarray}
f'(x) = -\frac{10}{3}f(x) + \frac{4\sqrt{6}}{x^{19/3}}
\sum_{n=1}^\infty e^{-b_n/x^2} b_n^{5/3}\left(U(-5/6,4/3,b_n/x^2) -
\frac{5}{6}U(1/6,7/3,b_n/x^2)  \right) \label{Airy_prime_dist}
\end{eqnarray}

Finally,
using $\sqrt{12} w_L = \sigma_1 L^{1/2} + {\cal O}(L^{-3/2})$ from Eq. (\ref{width}),
we can express the scaling in terms of the width $w_L$ instead of $L^{1/2}$. The first
two terms are given by
\begin{eqnarray}
P(h_m,L) = \frac{1}{\sqrt{12}\, w_L} f\left(\frac{h_m}{\sqrt{12}\,
 w_L}\right) + \frac{1}{\sqrt 2\, \sigma_1\, L} f'\left(\frac{h_m}{\sqrt{12}\,
 w_L}\right) + \dots
\label{final_P} 
\end{eqnarray}
In particular, from Eq. (\ref{final_P}), one
obtains the mean value $E[h_m]$ as
\begin{eqnarray}
E[h_m] = \sigma_1\, \sqrt{L} - \frac{\sigma_1}{\sqrt{2}} + {\cal O}(L^{-1/2})
\label{mean1}
\end{eqnarray}
The leading correction to the asymptotic $L^{1/2}$ behavior of the expected 
MRH turns out to
be a negative constant, as in the case of the expected maximum of an ordinary random walker
treated recently in Refs.~\cite{flajolet_correc_scal,CM,CMZ}.

The exact solution presented in this section proves that indeed the leading
scaling function of the MRH pdf in the SOS model with $p=1$ 
is given by the Airy distribution function, in agreement with our
general argument presented in section II. In the next section we
present numerical simulations for other values of $p$ to provide
further support in favour of the universality of the Airy distribution function.
In addition, the exact solution for $p=1$ also provides the leading correction
to scaling function $f_1(x)$ exactly. We will see in the next section
that simulations for other values of $p$ indicate that even this leading
correction to scaling function $f_1(x)$ is also universal, {\it i.e.}
independent 
of $p$.

\section{Numerical Simulations}

In order to generate numerically height profiles $H_i$ (with pbc) according to
the Boltzmann-weight associated to ${\cal H}_p$ in Eq. (\ref{def_hamiltonian})
we use the algorithm proposed in \cite{MDG} which
circumvents the use of slow Monte-Carlo algorithm with relaxation time $\sim L^2$. 
It relies upon the fact that the variables
$\{H_i\}_{0\leq i \leq L}$, given
that $(H_{i+1}-H_i)$ are weakly correlated random variables {\it with} their
total sum equal to zero due to pbc, can
be viewed as a Brownian bridge. To 
generate such a Brownian bridge, we first generate an ordinary random walk
sequence
${\cal W}_i$ such that ${\cal W}_0 = 0$ 
and ${\cal W}_{i} = {\cal W}_{i-1} + \eta_i$,
where $\eta_i$'s 
are i.i.d. random variables drawn from ${\rm Prob}(\eta=x)=g_p(x)$ given in Eq. (\ref{noisedist}). 
Next we express the $H_i$
variables in terms of the ${\cal W}_i$ variables by the relation
$H_i = {\cal W}_i - (i/L){\cal W}_L$, such that they satisfy the same recursion
relation as the ${\cal W}_i$'s, but this form ensures
that the sequence of $H_i$'s is periodic with period $L$ and hence automatically
enforces the pbc $H_0 = H_L$. One can show that this procdure yields
the statistical weight for a configuration of $H_i$'s, at least for large $L$
(except for the $p=2$ case where it is exact for all $L$).
Once the actual height sequence 
$H_i$ is generated,
the relative height profile $h_i$ is then obtained from the relation
$h_i= H_i - \sum_i H_i/L$. For every sample of $h_i$ sequence, we measure the maximum
$h_m$ and then obtain its histogram $P(h_m,L)$ from a total of
$2\cdot 10^6$ configurations of the height profile. 
In addition, we also compute the width $w_L$ from $w_L^2=\sum_{i} \langle h_i^2\rangle/L$.

\begin{figure}[h]
\begin{minipage}{0.43 \linewidth}
\begin{center}
\includegraphics[width=\linewidth]{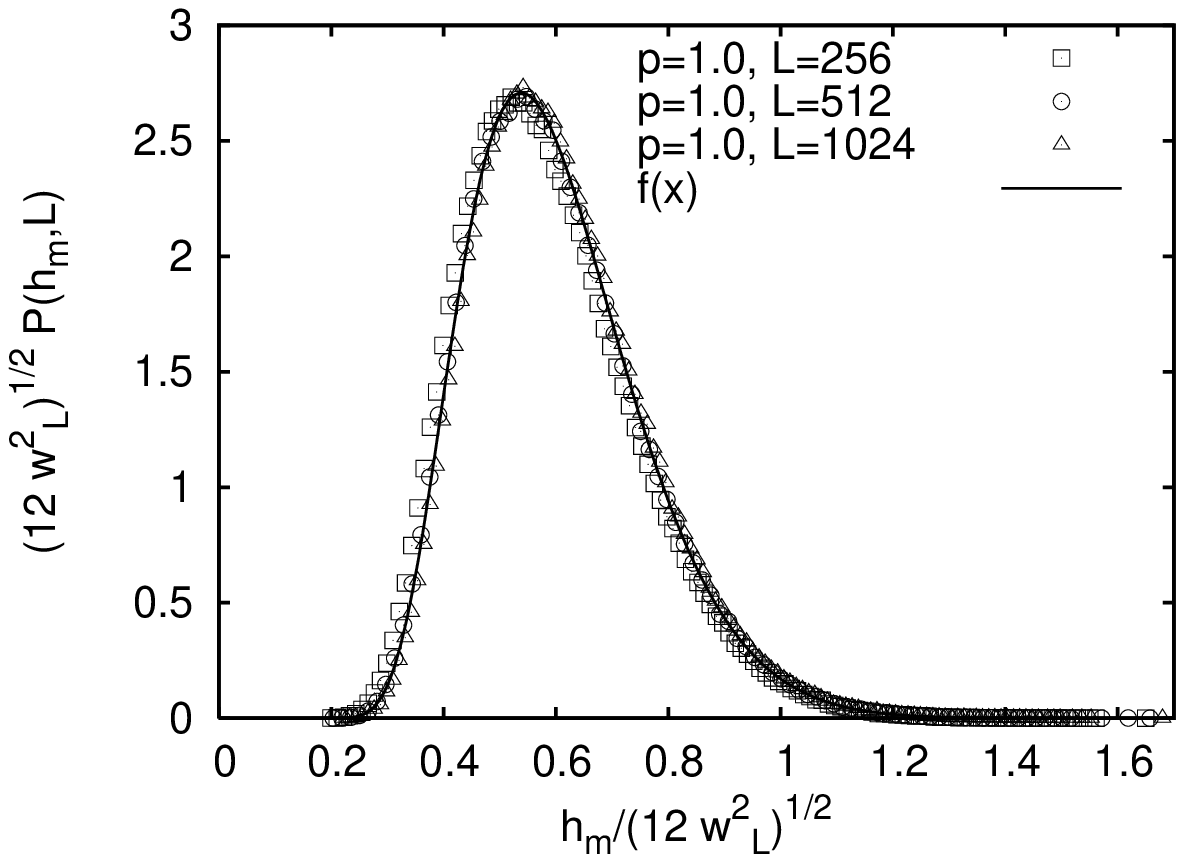}
\caption{$\sqrt{12 w^2_L} P(h_m,L)$ as a function of $h_m/\sqrt{12
    w^2_L}$ for different system sizes $L=256,512$ 
and $1024$ for $p=1$. The solid line is the Airy-distribution function
    (\ref{Airy_dist}) evaluated using Mathematica. There is no fitting
    parameter.} \label{Fig1}
\end{center}
\end{minipage}\hfill
\begin{minipage}{0.43 \linewidth}
\vspace*{-0.5cm}%
\includegraphics[width=\linewidth]{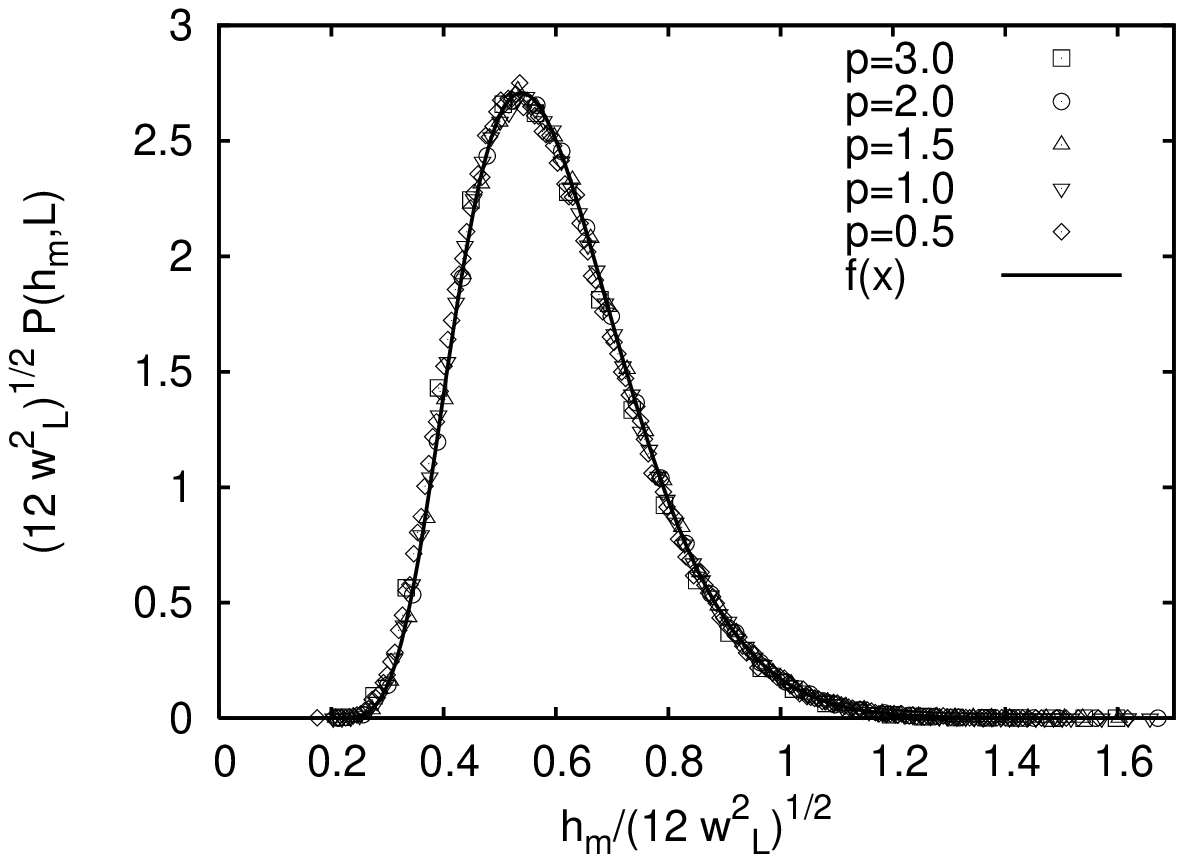}
\caption{$\sqrt{12 w^2_L} P(h_m,L)$ as a function of $h_m/\sqrt{12
    w^2_L}$ for $p=0.5,...,3$ and $L=512$. The solid line is the
    Airy-distribution function 
    (\ref{Airy_dist}) evaluated using Mathematica. There is no fitting
    parameter. 
}\label{Fig2}
\end{minipage}
\end{figure}

In Fig. \ref{Fig1}, we show a plot of $\sqrt{12 w^2_L}\, P(h_m,L)$ for
$p = 1$ as a function of the rescaled
variable $h_m/\sqrt{12 w^2_L}$ for different system sizes $L=256,512$
and $1024$. The good collapse shows a good agreement with our
analytical approach for $p=1$. Interestingly, our numerical
simulations show that the leading behavior of $P(h_m,L)$ as a function
of $h_m/\sqrt{12 w^2_L}$ (\ref{final_P}) is in fact
independent of $p$ and given by the Airy-distribution function
$f(x)$ (\ref{Airy_dist}). This is depicted in Fig. \ref{Fig2} where we
show a plot of $\sqrt{12 w^2_L} P(h_m,L)$ for different values of
$p=0.5,1,1.5,2$ and $3$ and fixed $L$ (for $p$ fixed and different
values of $L$ one obtains similar results as for $p=1$ in Fig. \ref{Fig1}).

\begin{figure}[h]
\begin{minipage}{0.43 \linewidth}
\begin{center}
\vspace*{-0.2cm}%
\includegraphics[width=\linewidth]{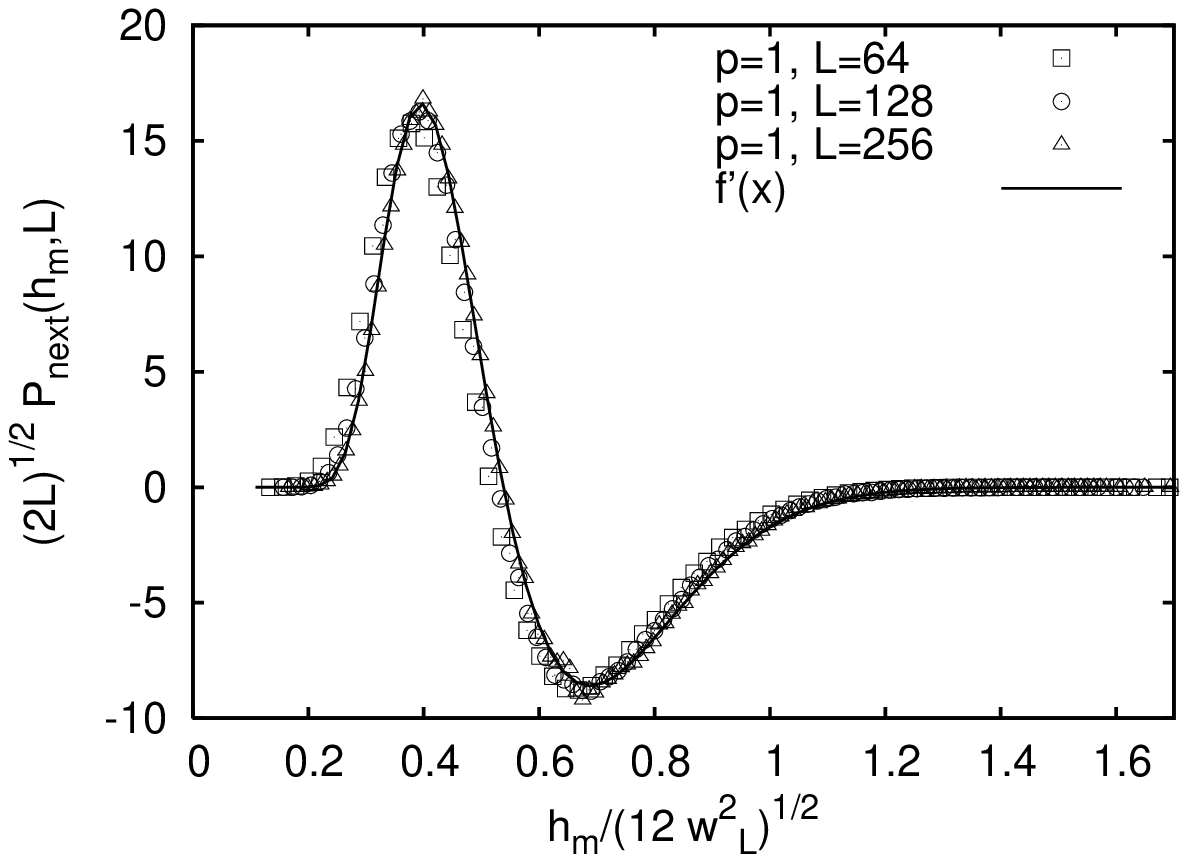}
\caption{$(2L)^{1/2}P_{\text{next}}(h_m,L)$
as a function of $h_m/\sqrt{12 w^2_L}$ for $p=1$ and different system
sizes $L=64,128, 256$. The solid line is the derivative of the
Airy-distribution function 
    (\ref{Airy_prime_dist}) evaluated using Mathematica. There is no
fitting parameter.} \label{Fig3} 
\end{center}
\end{minipage}\hfill
\begin{minipage}{0.43 \linewidth}
\includegraphics[width=\linewidth]{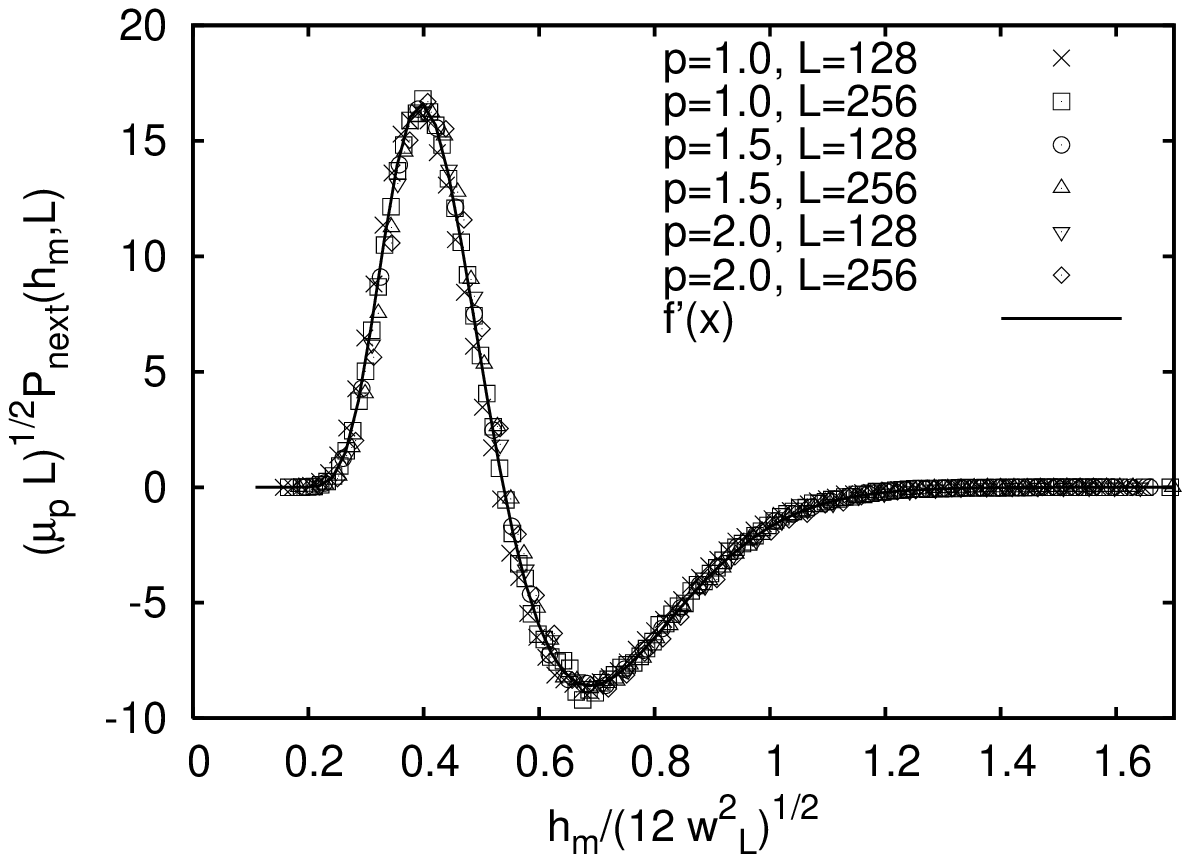}
\caption{$(2L)^{1/2}P_{\text{next}}(h_m,L)$
as a function of $h_m/\sqrt{12 w^2_L}$ for different system sizes
$L=128, 256$ and for $p=1,1.5,2$, with $\mu_1=2$(exact), $\mu_{1.5} =
3.8(1)$ and $\mu_2 = 6.9(1)$. The solid line is the derivative of the
Airy-distribution function 
    (\ref{Airy_prime_dist}) evaluated using Mathematica.}\label{Fig4}
\end{minipage}
\end{figure}

We have also investigated numerically the corrections to scaling. To
do so, we introduce the quantity $P_{\text{next}}(h_m,L)$ 
\begin{eqnarray}
P_{\text{next}}(h_m,L) = \sqrt{12 w^2_L} P(h_m,L) - f(h_m/\sqrt{12
  w^2_L}) \label{def_Pnext}
\end{eqnarray}
For $p=1$, according to our analytical calculations, the leading
behavior of $P_{\text{next}}(h_m,L)$ is given by
$P_{\text{next}}(h_m,L)~\sim~(2L)^{-1/2} f'(h_m/(12 w^2_L)^{1/2})$. In
Fig. \ref{Fig3}, we show a plot of $(2L)^{1/2}P_{\text{next}}(h_m,L)$
as a function of the rescaled variable $h_m/\sqrt{12 w^2_L}$, for
$p=1$, and different system sizes $L=64,128$ and $256$. Here also, one
obtains a good agreement with our analytical calculation
(\ref{Airy_prime_dist}). Finally, we have also computed numerically
the corrections to scaling (\ref{def_Pnext}) for different values of
$p$. Our data (see Fig. \ref{Fig4}) suggests that the leading
correction to scaling, for any $p$, takes the following form
\begin{eqnarray}
P_{\text{next}}(h_m,L) \sim \frac{1}{\sqrt{\mu_p L}} f'(h_m/\sqrt{12
  w^2_L}) \label{P_next_scal}
\end{eqnarray}
where $\mu_1 = 2$. 
In Fig. \ref{Fig4}, we show a plot of $(\mu_p
L)^{1/2}P_{\text{next}}(h_m,L)$ 
as a function of the rescaled variable $h_m/\sqrt{12 w^2_L}$ for 
$L=128, 256$ and $p=1,1.5$ and $2$. In this figure, $\mu_{1.5} = 3.8(1)$ and
$\mu_2 = 6.9(1)$ are estimated to obtain the best collapse with
$f'(x)$. The good data collapse which we obtain is indeed in good
agreement with the behavior in Eq.~(\ref{P_next_scal}). And therefore
the corrections to scaling in Eq. (\ref{final_P}) are also independent
of $p$, up to a non universal amplitude $\mu_p$.

\section{Conclusion}

In this paper, we have studied the distribution of the maximal relative height $h_m$
in a class of one dimensional solid-on-solid lattice models defined by the Hamiltonian
in Eq. (\ref{Hamilg}). We have provided a simple central limit argument 
to show that for arbitrary symmetric $g(x)$ with a finite second moment $\sigma^2$,
the asymptotic MRH distribution $P(h_m,L)$ has a scaling form, 
$P(h_m,L)\approx (1/{\sqrt{12} w_L}) f\left(h_m/{\sqrt{12} w_L}\right)$
where the width $w_L^2\approx {\sigma^2 L}/12$ for large $L$ depends only on
the variance $\sigma^2$, but the scaling function $f(x)$ is universal (independent
of the details of the function $g(x)$) and is given by the Airy distibution function
in Eq. (\ref{Airy_dist}). This argument is supported by exact calculation
in the case of the Hamiltonian in Eq. (\ref{def_hamiltonian}) with $p=1$
and also by numerical simulations of the model in Eq. (\ref{def_hamiltonian}) for
other values of $p$. Moreover, we have shown that even the leading correction to
scaling function is also universal (up to an overall nonuniversal amplitude) and is given by the 
derivative of the Airy distribution function.   

There remain several open questions for future studies. In this paper we have studied symmetric
$g(x)$ functions with a finite second moment $\sigma^2$. We expect that the universality
class of the asymptotic MRH distribution will change if $\sigma^2$ is not finite.
For example, if $g(x)\sim x^{-(1+\delta)}$ for large $x$ with $0\le \delta\le 2$ that
corresponds to the L\'evy walk of the interface, it would be interesting to compute
the MRH distribution which we expect will depend on $\delta$. Another interesting
case would be to study the MRH distribution with asymmetric $g(x)$ 
that corresponds to interfaces 
with a drift.  

In this paper we have focused on very simple SOS Hamiltonians that have only 
`kinetic' or `interaction' part, but no onsite potential. There are important applications 
of the SOS Hamiltonians with an onsite attractive potential in wetting 
phenomena~\cite{Forgacs,discrete}
as well as in understanding the thermal denaturation of DNA molecules~\cite{DNA}.
In this case one considers a Hamiltonian of the form
\begin{equation}
{\cal H} =  \sum_i \left[F\left(|H_{i}-H_{i+1}|\right)+ U(H_i)\right]\,,
\label{Hamilg1}
\end{equation}
where the heights $H_i \in  ]-\infty, +\infty[$, $F(x)=-\ln[g(x)]$ as in Eq. (\ref{Hamilg})
and $U(H_i)$ 
represents a finite attractive 
onsite potential. One can ask: what is the MRH distribution in these models? In this case,
both the heights $H_i$'s as well as the relative heights $h_i$'s will be bounded variables
in equilibrium. This follows from the transfer matrix analysis in section IV and the fact
that any localized potential (far from the boundary) in one dimension
will always give rise to a bound state. Consequently, the correlation function between the 
heights $h_i$'s will decay exponentially with distance. Thus, the MRH distribution
in this case corresponds to calculating the distribution of the maximum of a
set of `weakly' correlated random variables. By considering `blocks' of
size of the correlation length, it is then easy to see that asymptotically
(when the system size $L$ is much larger than the corelation length) the MRH
distribution will follow~\cite{satya_mrh2}, after a suitable rescaling, the Gumbel law of the extreme
of uncorrelated random variables~\cite{EVS}. 

On the other hand, if one considers the heights $H_i$'s to be positive
variables, $H_i\ge 0$, {\it i.e.}
one is considering a semi-infinite system, then the presence of a localized potential
near the boundary $H=0$ leads to a roughening phase transition at a critical
temperature $T_c$~\cite{discrete,burkhardt}. Below $T_c$, the interface
is bound, {\it i.e.} localized near the surface at $H=0$ (with a finite $L$ independent width
for large $L$), but above $T_c$ the
system becomes unbound, {\it i.e.} rough with width $w_L\sim L^{1/2}$. Consequently, we would
expect that the asymptotic MRH distribution will have a Gumbel form below
$T_c$, but for $T>T_c$ it will change to the Airy distribution function described in this paper.
In short, the bound phase will be characterized by the Gumbel form and the rough phase
by the Airy distribution function. It would be interesting to compute how this
transition in the MRH distribution takes place across the critical temperature.
Finally it would be interesting to compute the MRH distribution in the lattice SOS models
in higher dimensions.

\acknowledgments

GS acknowledges the financial support provided
through the European Community's Human Potential Program 
under contract HPRN-CT-2002-00307, DYGLAGEMEM.

\appendix

\section{ Exact calculation of the width $w_L$ in the SOS model}

We wish to compute the width $w_L$ defined via
\begin{equation}
w_L^2=\frac{1}{L}\sum_{i=1}^L \langle h_i^2 \rangle 
\label{w1}
\end{equation}
where $h_i$'s are the relative heights 
\begin{equation}
h_i=H_i-\frac{1}{L}\sum_i H_i 
\label{rh1}
\end{equation}
and $H_i$'s are the absolute heights in the SOS model defined by Eq. (\ref{Hamilg}).
Substituting Eq. (\ref{rh1}) in (\ref{w1}) one gets
\begin{equation}
w_L^2= \frac{L-1}{L^2}\sum_{i=1}^L \langle H_i^2\rangle -\frac{2}{L^2}\sum_{i<j} \langle H_i 
H_j\rangle.
\label{a1}
\end{equation}

The absolute height profile $H_i$'s satisfy the recursion relation
\begin{equation}
H_{i}= H_{i-1} + \xi_i\,.
\label{a2}
\end{equation}
It follows from Eq. (\ref{Hamilg}) that
the set of random numbers $\{\xi_i\}$ with $i=1,2,\ldots, L$ are distributed
according to the joint law
\begin{equation}
P\left[\{\xi_i\}\right]= N_L\, g(\xi_1)\,g(\xi_2)\,\ldots g(\xi_L)\, \delta\left(\sum_{i=1}^L 
\xi_i\right)
\label{a3}
\end{equation}
where $N_L$ is such that the joint distribution is normalized and $g(x)$ is 
a symmetric, normalized pdf with a finite second moment
$\sigma^2$, but otherwise arbitrary. The delta 
function 
ensures that the path is periodic, $H_0=H_L$. 
Thus, the variables
$\xi_i$'s are only `weakly' correlated due to the global constraint $\sum_i\xi_i=0$.
Since we are only interested in relative heights
in Eq. (\ref{rh1}), we can, without any loss of generality, set $H_0=H_L=0$.
Thus, from Eq. (\ref{a2}) we have
\begin{equation}
H_i = \sum_{k=1}^{i} \xi_k \,.
\label{a4}
\end{equation}

To calculate the correlation function $\langle H_i H_j\rangle$ as required in Eq. (\ref{a1})
we need to compute the correlation function $\langle \xi_j\xi_k\rangle$ of the noise
variables. This is easy to do. From the constraint $\sum_i \xi_i=0$, it follows
that $\langle \xi_k \sum_i \xi_i \rangle =0$ for all $k$. Using the isotropic
property ({\it i.e.} $\langle \xi_k\xi_j\rangle$ is the same for any pair of $(k,j)$
as long as $k\ne j$), it follows that
\begin{equation}
\langle \xi_k\xi_j\rangle = -\frac{\langle \xi_i^2\rangle}{L-1}
\label{a5}
\end{equation}
where the onsite variance $\langle
\xi_i^2\rangle$ is independent of $i$ and will be computed later.
Eqs. (\ref{a5}) and (\ref{a4}) give the required height correlations
\begin{eqnarray}
\langle H_i H_j\rangle &=& \frac{\langle\xi_i^2\rangle}{L-1}\, i(L-j)\quad {\rm for}
\quad i\le j \nonumber \\
&=& \frac{\langle\xi_i^2\rangle}{L-1}\, j(L-i) \quad {\rm for}\quad i\ge j \,.
\label{a6}
\end{eqnarray}
We now substitute this result in Eq. (\ref{a1}). We need to use one identity, namely,
$\sum_{i<j} i(L-j)= L(L+1)(L-2)(L-1)/24$.  Subsequently, a straightforward 
algebra gives 
\begin{equation}
w_L^2 = \frac{L+1}{12}\, \langle \xi_i^2\rangle\,.
\label{a7}
\end{equation}

The only remaining task is to compute the onsite variance $\langle \xi_i^2\rangle$
where the set $\{\xi_i\}$'s is drawn from the joint distribution in Eq. (\ref{a3}).
Let $P_L(x)= {\rm Prob}(\xi_i=x)$ be the `single site' probability density of the variable
$\xi_i$, obtained by keeping $\xi_i=x$ fixed and integrating the joint distribution in Eq. (\ref{a3})
over the rest $(L-1)$ variables. Evidently, it follows from the isotropic property that
$P_L(x)$ is independent of $i$. By integrating over the $(L-1)$ variables in
Eq. (\ref{a3}) one gets
\begin{equation}
P_L(x)= N_L\, g(x)\, V_{L-1}(-x)
\label{a8}
\end{equation}
where $V_n(x)$ is the mutiple integral
\begin{eqnarray}
V_n(x)&=&\int g(x_1)\, g(x_2)\, \ldots g(x_n)\, \delta\left(\sum_{i=1}^n x_i-x\right) \nonumber \\
&=& \int g(x_1)\, V_{n-1}(x-x_1)\, dx_1\,.
\label{a9}
\end{eqnarray}
The recursion relation satisfied by $V_n(x)$ above starts with the initial condition
$V_0(x) =\delta(x)$ and has a simple convolution form. Hence it can be solved exactly by 
the Fourier transform method. Let, ${\tilde V_n}(k)=\int_{-\infty}^{\infty} V_n(x) e^{ikx} dx$
and ${\tilde g}(k) =\int_{-\infty}^{\infty} g(x) e^{ikx} dx$. Taking the Fourier transform
of Eq. (\ref{a9}) gives
\begin{equation}
{\tilde V_n}(k)= {\tilde g}(k)\, {\tilde V_{n-1}}(k)= \left[{\tilde g}(k)\right]^n
\label{a10}
\end{equation}   
where we have used ${\tilde V_0}(k)=1$.
We invert the Fourier transform and substitute in Eq. (\ref{a8}) to get
\begin{equation}
P_L(x) = N_L\, g(x)\, \int_{-\infty}^{\infty} \left[{\tilde g}(k)\right]^{L-1} 
e^{-ikx} dk\,.
\label{a11}
\end{equation}
The constant $N_L$ is fixed by demanding $\int_{-\infty}^{\infty} P_L(x)dx=1$. 
Using the symmety ${\tilde g}(k)={\tilde g}(-k)$ we then get
\begin{equation}
P_L(x) = g(x)\, \frac{\int_{-\infty}^{\infty} \left[{\tilde g}(k)\right]^{L-1} e^{-ikx} dk}{
\int_{-\infty}^{\infty} \left[{\tilde g}(k)\right]^{L} dk}\,.
\label{a12}
\end{equation}
We are now ready to compute the onsite variance, $\langle \xi_i^2\rangle =\int P_L(x) x^2 dx$.
We get from Eq. (\ref{a12})
\begin{equation}
\langle \xi_i^2\rangle = - \frac{\int_{-\infty}^{\infty}\left[{\tilde g}(k)\right]^{L-1} {\tilde 
g}''(k) dk}
{\int_{-\infty}^{\infty} \left[{\tilde g}(k)\right]^{L} dk}
\label{a13}
\end{equation}
where ${\tilde g}''(k) = d^2 {\tilde g}/dk^2$.
The result in Eq. (\ref{a13}) can be further simplied via integration by parts. Substituting this 
result in Eq. (\ref{a7}) we get our final exact result, valid for all $L$,
\begin{equation}
w_L^2=\frac{L^2-1}{12}\,\Delta_L \quad, \quad
\Delta_L=\frac{\int_{-\infty}^{\infty} [{\tilde g}(k)]^{L-2}\,[{\tilde g}'(k)]^2\, dk}
{\int_{-\infty}^{\infty} [{\tilde g}(k)]^{L}\, dk}
\label{widtha}
\end{equation}
where ${\tilde g}'(k)= d{\tilde g}/dk$.

Let us now consider a few special cases. For the Gaussian case, {\it i.e.} for $p=2$ in Eq. 
(\ref{def_hamiltonian}), we have ${\tilde g}(k)=\exp[-\sigma^2k^2/2]$.
From Eq. (\ref{widtha}) we get for all $L\ge 1$
\begin{equation}
w_L^2 = \frac{L^2-1}{12 L} \sigma^2\,.
\label{a14}
\end{equation} 
For the $p=1$ case in Eq.
(\ref{def_hamiltonian}), we have $g(x) = K \exp[-K|x|]/2$. The Fourier transform gives
${\tilde g}(k) = K^2/(K^2 + k^2)$. Thus in this case, $\sigma^2=-{\tilde g}''(0)= 2/K^2$. 
Substituting in Eq. (\ref{widtha}) and performing 
the integral we get, for all $L\ge 1$
\begin{equation}
w_L^2= \frac{(L^2-1)\,(2L-1)}{24\, L\,(L+1)}\sigma^2\,.
\label{a15}
\end{equation}

In general for arbitrary symmetric $g(x)$ with a finite second moment $\sigma^2$, one
gets from Eq. (\ref{widtha}) the following large $L$ behavior,  
\begin{equation}
w_L^2\approx \sigma^2 \, \frac{L}{12}\, .
\label{a16}
\end{equation}
This can be seen from the fact that for large $L$, the Fourier integrals in Eq. (\ref{widtha}) 
will be dominated by the contributions from the small $k$ limit. For small
$k$, ${\tilde g}(k) \approx 1- \sigma^2k^2/2 +...\approx \exp[-\sigma^2k^2/2]$. Substituting this 
result
in Eq. (\ref{widtha}) and performing the integrals one arrives at the asymptotic behavior in
Eq. (\ref{a16}).


\begin{thebibliography}{99}

\bibitem{TempPers} J. Krug et. al. Phys. Rev. E, 56, 2702 (1997); 
H. Kallabis and J. Krug, Europhys. Lett. {\bf 45}, 20 (1999);
S.N. Majumdar, Current Science, {\bf 77}, 370 (1999). 
J. Krug, Physica A {\bf 340}, 647 (2004);
M. Constantin et. al. Phys. Rev. E, {\bf 69}, 061608 (2004);
C. Dasgupta et. al. {\it ibid}, {\bf 69}, 022101 (2004); S.N. Majumdar
and D. Das, {\it ibid}, {\bf 71}, 036129 (2005).

\bibitem{SpatPers} S. N. Majumdar and A. J. Bray, Phys. Rev. Lett. {\bf 86},
3700 (2001); M. Constantin, S. Das Sarma, and C. Dasgupta, Phys. Rev.
E {\bf 69}, 051603 (2004).

\bibitem{MDG} S. N. Majumdar and C. Dasgupta, cond-mat/0509109
(to appear in Phys. Rev. E (2006)).

\bibitem{EW} S.F. Edwards and D.R. Wilkinson, 
Proc. R. Soc. London A {\bf 381}, 17 (1982).

\bibitem{KPZ} M. Kardar, G. Parisi, and Y.-C. Zhang,
  Phys. Rev. Lett. {\bf 56}, 889 (1986). 

\bibitem{Exp} D. B. Dougherty et. al. Phys. Rev. Lett. {\bf 89},
136102 (2002); D. B. Dougherty et. al. Surf. Sci. {\bf 527}, L213 (2003).
O.~Bondarchuk et. al. Phys. Rev. B {\bf 71}, 045426 (2005);
D. B. Dougherty et. al. 
Phys. Rev. E {\bf 71}, 021602 (2005).  

\bibitem{GHPZ} G. Gyorgyi, P.C.W. Holdsworth, B. Portelli, and Z. Racz, Phys. Rev. E,
{\bf 68}, 056116 (2003).

\bibitem{RCPS} S. Raychaudhuri, M. Cranston, C. Przybyla, and Y. Shapir, Phys. Rev.
Lett. {\bf 87}, 136101 (2001).

\bibitem{satya_mrh}
S.~N.~Majumdar and A.~Comtet, Phys. Rev. Lett., {\bf 92}, 225501 (2004).

\bibitem{satya_mrh2} S.~N.~Majumdar and A.~Comtet, J. Stat. Phys, {\bf 119} 
777 (2005).

\bibitem{GK} H. Guclu and G. Korniss, Phys. Rev. E {\bf 69}, 065104(R) (2004);
Fluctuation and Noise Letters, {\bf 5} (1), L43 (2005).

\bibitem{Lee} D.S. Lee, Phys. Rev. Lett. {\bf 95}, 150601 (2005). 

\bibitem{EVS} E. J. Gumbel, {\em Statistics of Extremes} (Columbia University
Press, New York, 1958); S. Coles, {\em An Introduction to Statistical Modeling of Extreme 
Values}, Springer Series in Statistics (Springer-Verlag, London, 2001).

\bibitem{corrEVS} S.N. Majumdar and P.L. Krapivsky, Physica A {\bf 318}, 161 (2003);


\bibitem{HZ} T. Halpin-Healy and Y.-C. Zhang, Phys. Rep. {\bf 254}, 215 
(1995).

\bibitem{bart} N. C. Bartelt, J. L. Golding, T. L. Einstein, and E. D.
Williams, Surf. Sci. {\bf 273}, 252 (1992).

\bibitem{abramowitz}
M. Abramowitz and I.A. Stegun in {\it Handbook of Mathematical
  Functions} (Dover, New York, 1973).

\bibitem{FPV} P. Flajolet, P. Poblete, 
and A. Viola, Algorithmica, {\bf 22}, 490 (1998).

\bibitem{BF} For a recent review on the Airy distribution function and its
diverse applications see S. N. Majumdar, cond-mat/0510064 (to appear
in Current Science, 2005).

\bibitem{excursion} D.A. Darling, Ann. Probab. {\bf 11}, 803 (1983); G. Louchard,
J. Appl. Prob. {\bf 21}, 479 (1984).

\bibitem{takacs_invert} L.~Takacs, J. Appl. Prob. {\bf 32}, 375 (1995).

\bibitem{Forgacs} G. Forgacs, R. Lipowsky, and Th. M. Nieuwenhuizen, in {\em Phase
Transitions and Critical Phenomena} ed. by C. Domb and J.L. Lebowitz (Academic Press, 
London, 1991), vol 14, 136 (1991).

\bibitem{discrete} D.B. Abraham, Phys. Rev. Lett. {\bf 44}, 1165 (1980); S.T. Chui and 
J.D. Weeks, Phys. Rev. B {\bf 23}, 2438 (1981); J.M.J. van Leeuwen and H.J. Hilhorst, 
Physica , {\bf 107 A}, 319 (1981).

\bibitem{burkhardt} T.W. Burkhardt, J. Phys. A {\bf 14}, L63 (1981).  

\bibitem{sinai} Y.G. Sinai, Theor. Math. Phys. {\bf 90}, 219 (1992).

\bibitem{burkhardt_jointprob}
T~.W.~Burkhardt, J. Phys. A, {\bf 26}, L1157-1162 (1993).

\bibitem{flajolet_correc_scal}
E.G.~Coffman, P.~Flajolet, L.~Flato and M.~Hofri, {Proba. in
  Engineering and Informational Sciences}, {\bf 12}, 373 (1998).

\bibitem{CM}
A.~Comtet and S.N. Majumdar, {J. Stat. Mech. Theor. Exp.}, {\bf 06},
P06013, (2005).

\bibitem{CMZ}
S.N. Majumdar, A. Comtet, and R.M. Ziff, cond-mat/0509613, to appear in J. Stat. Phys. (2006).

\bibitem{DNA} M. Peyrard and A.R. Bishop, Phys. Rev. Lett. {\bf 62}, 2755 (1989); T. Dauxois,
M. Peyrard, and A.R. Bishop, Phys. Rev. E {\bf 47}, 684 (1993); S. Ares and A. Sanchez, 
cond-mat/0511532.

\end{thebibliography}
\end{document}